\documentclass[amsmath,amssymb,prx,floatfix,twocolumn,superscriptaddress]{revtex4-2}

\usepackage{graphicx}
\usepackage{dcolumn}
\usepackage{bm}

\begin{document}

\title{Weighted network motifs as random walk patterns}

\author{Francesco Picciolo}
\affiliation{Department of Physical Sciences, Earth and Environment, University of Siena, 53100 Siena, Italy}

 \author{Franco Ruzzenenti}
\affiliation{Integrated Research on Energy, Environment and Society, Faculty of Science and Engineering, University of Groningen, Nijenborgh 7, 9747AG Groningen, The Netherlands}

\author{Petter Holme}
\affiliation{Tokyo Tech World Hub Research Initiative (WRHI), Institute of Innovative Research, Tokyo Institute of Technology, Tokyo 152-8550, Japan}
\author{Rossana Mastrandrea}
\affiliation{IMT School for Advanced Studies, Lucca, piazza S. Ponziano 6, 55100 Lucca, Italy}

\begin{abstract}
Over the last two decades, network theory has shown to be a fruitful paradigm in understanding the organization and functioning of real-world complex systems. One technique helpful to this endeavor is identifying functionally influential subgraphs, shedding light on underlying evolutionary processes. Such overrepresented subgraphs, \textit{motifs}, have received much attention in simple networks, where edges are either on or off. However, for weighted networks, motif analysis is still undeveloped. Here, we proposed a novel methodology---based on a random walker taking a fixed maximum number of steps---to study weighted motifs of limited size. We introduce a sink node to balance the network and allow the detection of configurations within an \textit{a priori} fixed number of steps for the random walker. We applied this approach to different real networks and selected a specific null model based on maximum-entropy to test the significance of weighted motifs occurrence. We found that identified similarities enable the classifications of systems according to functioning mechanisms associated with specific configurations: economic networks exhibit close patterns while differentiating from ecological systems without any \textit{a priori} assumption.
\end{abstract}

\maketitle

\section{Introduction}

\subsection{An overview of network motifs}

In recent years, network theory has become a core of the interdisciplinary study of complex systems. Many global features of the network emerge as a result of the underlying local properties. Therefore, the analysis of network patterns can reveal the system's structural organization and functioning mechanism and possibly shed light on the process of originating specific configurations during network evolution. The identification of network \textit{motifs} has received increasing attention starting from the seminal work by Milo \textit{et al.}~\cite{milo2002network} (although there were precursors already 30 years before~\cite{holland_leinhardt}). The authors argued that the frequency of small subgraphs (motifs) could characterize complex networks. Such ``simple building blocks'' could connect the way the systems operate to its network structure and further its evolution~\cite{shen2002network,mangan2003structure,alon2007network}.

Mathematically speaking, a motif $M$ is itself a graph---usually, but not necessarily, directed. An occurrence of this motif $M$ in a larger graph $G$ means that there is a relabeling of $G$'s nodes such that the edges involving nodes in $M$ are exactly the same as in $M$. As the number of nodes increases, the number of possible patterns increases super-exponentially, making their counting very complex. For this reason, much research focus, recently, has been on developing fast algorithms~\cite{ugander2013subgraph,jha2015path,saha2015finding,wang2015minfer}. 

The significance of motifs occurrence in a system can be tested comparing with a null model: a randomized version of the original network where one or more local properties are kept fix (degree, strength) and the wiring of the links is randomized accordingly. 
The choice of the null model represents a fundamental step, often not carefully treated, as the selection of the  constraints could  affect the significance of the network properties under study. Indeed, the use of a null model has the goal to wash away statistical effect simply due for example to network size~\cite{artzy2004comment,thusresponse}.

\subsection{Weighted network motifs}

By associating weights to the edges, one can encode more relevant information about the system than is possible in a simple graph of binary edges~\cite{newman2004analysis,barrat2004architecture,serrano2009extracting,mastrandrea2017organization,amador2018s}. It comes thus natural to search for an extension of the motif concept to the weighted case. However, only a few works have taken this direction, using one of two ideas. First, some authors study the configurations of weights on the binary motifs. Second, some works refined the concept of the motif itself via random walks on the network. While the first line of research tends to compare results from the weighted and unweighted methods evaluating to what extent the introduction of edge-values alter the significance profile of binary motifs; the random walk procedure has primarily been used as an intermediate step towards the identification of network communities.

The association of weights to binary motifs is not straightforward as it strictly depends on the kind of network. One should consider the nature of weights disentangling between qualitative and quantitative edge-links, discretized or continuous weights, positive and negative values, correlation coefficients (weights $\in [0,1]$), and so on. Ref.~\cite{onnela2005intensity} described this type of weighted motifs as a ``set of topologically equivalent subgraphs of a network'' and introduced the concepts of \textit{intensity} and \textit{coherence} as natural extension of motifs in the unweighted case. The intensity is computed as the geometric mean of the edge weights forming a specific binary pattern, while its ratio to the corresponding arithmetic mean represents the coherence. They used the metabolic directed network of \textit{E. coli} to show how the introduction of weights can, in some cases, completely reverse the significance profiles of the corresponding underlying binary motif.

In 2012, Chobdar \textit{et al.}~\cite{choobdar2012motif}
associated to a binary motif a \textit{weight entropy} looking at the occurrence of a weighted subgraph as an event, such that its weight distribution has the form of a probability function. hey tested their approach on the network of co-authorship of publications from the University of Porto, ranging from 2003 to 2011.The weighted case resulted as good as the unweighted one, with the advantage of being less time consuming. 
Later, Chobdar \textit{et al.}~\cite{choobdar2015discovering} proposed an alternative way of defining the significance profile of motifs: they identified a subgraph as a motif when the edge weights followed a distribution significantly different from a random distribution. This approach was tested on the gene-coexpression network and revealed that the weighted motifs so defined are more biologically relevant than the corresponding simple graphs.

In the same spirit other works have been developed on the concept of \textit{frequent subgraphs} (with few extensions to the weighted case) especially in the engineering field~\cite{kuramochi2001frequent,kuramochi2004efficient,bringmann2008frequent,jia2009towards,Hong2014}. The concept is slightly different from that of network motifs, even if the subgraph definition is the same. These approaches aim to enumerate subgraphs of the projected simple graph of edges with weights exceeding a preset threshold. To speed up the computation, they use various pruning criteria. However, these results are out of the scope of this paper. Rather, we aim at characterizing complex systems by looking at the composition and significance profile of their weighted subpatterns.

\subsection{Random walks to measure motifs}

Using walks of different kinds is a common idea to reduce a network to a string of symbols. This enables many general tools like representational learning~\cite{node2vec} or data compression~\cite{rosvall2008maps}. It is also typical approach to defining motifs in temporal networks~\cite{Kovanen18070,paranjape2017motifs}. 
Another group of works used the concept of random walk to define significant network subgraphs both for binary and weighted networks. The basic idea consists of placing a walker on a random node and move it to neighbors according to some transition probabilities depending on node strengths or degrees. There is the possibility also to introduce a restart probability to control how far the walker can arrive from the starting node~\cite{can2005analysis,lovasz1993random,el2012biological}. The approach has been used to estimate the clustering coefficient of an unweighted network~\cite{hardiman2013estimating} and to identify subgraphs of sizes three, four, and five~\cite{bhuiyan2012guise,wang2014efficiently}. However, most of these studies have been developed in the engineering field and aimed to find the most efficient algorithm to identify network subgraphs via a random walk approach. Here, we are instead interested in the use of random walk to characterize and classify real networks. With this in mind, we can find techniques based on random walks to identify network communities. A natural extension of the study of relevant network subgraphs of size $\le 5$. Ref.~\cite{rosvall2008maps} used the probability flow of network random walks as a proxy for information flows in the real-world system. They decomposed the network into groups of nodes by compressing a description of the probability flow. Ref.~\cite{zlatic2010topologically} presented a method based on the topology of biased random walks to explore complex undirected networks and identify their organization in communities. 
The main ingredients of our approach are weighted unbalanced networks, a \textit{sink node}, and a random walker. We introduced a sink node to compensate for the excess of inward flows of nodes to balance the network. The presence of a sink node allows computing the frequency of paths of any possible length observable within a fixed number of steps of a random walker placed on an arbitrary node. In other terms, all subgraphs of dimension smaller than and equal to the maximum number of steps can be considered mutually exclusive events. I.e., their total occupation probability sums up to one. The link weights and their heterogeneous distribution for each set of nodes play key roles. In principle, we can use this approach to find the network composition in significant subgraphs of any size, just varying the maximum number of steps of the random walker. However, simply increasing the number of steps will induce a combinatorial explosion to the problem since the number of possible motifs increases fast.

In the remainder of this paper we will apply our methodology to different real networks to study their composition according to weighted motifs occurrences. We test their significance to identify ``weighted building blocks'' similarly to the unweighted case explored by Ref.~\cite{milo2002network}. Furthermore, we show some applications to specific patterns and temporal trends and tested a set of null models to evaluate how local constraints can affect such a network characterization.

\section{Preliminaries}

\subsection{A random walk with a sink node}

Let be $G = (V, E)$ a graph with $V$ and $E$, respectively, the set of nodes and edges, such that $|V| = N$ and $|E| = L$. Let us assume that $G$ is a directed and weighted network; $A = (a_{ij})_{1\le i,j \le N}$ and $W = (w_{ij})_{1\le i,j \le N}$ are, respectively, the associated binary and weighted adjacency matrices. We can compute the node in and out strengths as:
\begin{equation}
s_i^{\rm out} = \sum_{j=1}^N w_{ij} \quad s_i^{\rm in} = \sum_{j=1}^N w_{ji} .
\end{equation}
A weighted network is \textit{balanced} if $s_i^{\rm out} = s_i^{\rm in} \quad \forall i \in\{1,\dots,N\}$.  In most of real-world networks $\exists k \in \{1,\dots,N\}$ such that $s_k^{\rm out} \neq s_i^{\rm in}$. In general, it is possible to identify three kinds of nodes: (i) balanced nodes for which $s_i^{\rm out} = s_i^{\rm in}$; (ii) not balanced nodes such that $s_i^{\rm out} > s_i^{\rm in}$; and (iii) not balanced nodes such that $s_i^{\rm out} < s_i^{\rm in}$. We introduce a novel node, called \textit{sink} and labelled as the $N+1$th node, to balance nodes with excess of in-strength
\begin{equation}
w_{iN+1} = s_i^{\rm in}-s_i^{\rm out} , \quad \forall i \in \{ j \in V | s_j^{\rm out} < s_j^{\rm in}\}
\end{equation}

The sink is connected only with nodes of type (iii). The total nodes outgoing flows will be given by:
\begin{equation}
ss_i^{out}= 
\begin{cases} 
s_i^{\rm out} & \text{if} \quad s_i^{\rm out} \geq s_i^{\rm in}\\
s_i^{\rm out} + w_{iN+1} & \text{if} \quad s_i^{out} < s_i^{\rm in}
\end{cases}
\end{equation}

We assume that a random walker placed on a node---randomly chosen in the set $V$---is allowed to move along its edges according to the transition probability matrix
\begin{equation} 
M = (m_{ij})_{1\le i,j \le N} , \text{ with } m_{ij}= \frac{w_{ij}}{ss_i^{\rm out}}
\end{equation} where $m_{ij}$ represents the probability for the random walker to move from node $i$ to node $j$, $\forall i,j \in \{1,\dots,N\}$. 

In principle, the random walker can move for an indefinite number of steps, but this increases the time-consumption for simulations and the complexity of analytical computations. Furthermore, here we are interested in computing the occurrence of significant network relatively small subpatterns, therefore we consider three as a good compromise starting point for the maximum number of steps of the walker.

\begin{figure}
\centering{\includegraphics[width=0.7\columnwidth]{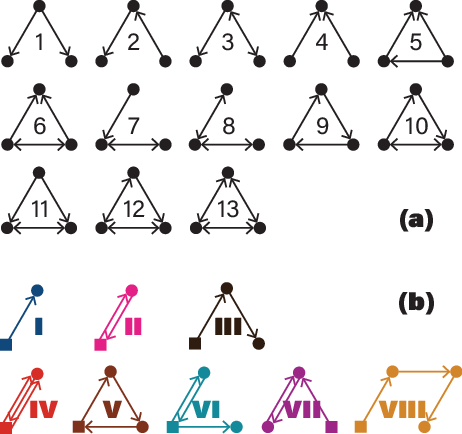}}
\caption{{\bf Binary and weighted motifs.} (a) Three-node binary motifs; (b) weighted subgraphs of sizes two, three, and four detectable by a random walker allowed to move for maximum three steps in a network with a sink node (square).}
\label{mot}
\end{figure}

In Fig.~\ref{mot}(a), we show all possible triadic directed binary motifs: subgraphs of size three with all possible combinations of directed links. Following Ref.~\cite{onnela2005intensity} one can extend the study to the weighted case, associating two novel measures---\textit{intensity} and \textit{coherence}---to the same 13 patterns. Differently, our approach consists of two ingredients: a sink node and a random walker moving for a limited number of steps, three in this case. Therefore, only some subgraphs of Fig.~\ref{mot}(a) are observable with this approach: 2, 3, 7 and 9. Indeed, by definition, a random walker starting from a random node can move along only one of its edges. Consequently, motifs 1 and 11 will never be described by his path. On the other hand, a target node cannot be reached from two different sources as in motifs 4, 5, 6, while we can explore motif 7 thanks to the presence of a reciprocated edge. Lastly, motifs number 8, 10, and 13 require more than three steps to be described. 

In Fig.~\ref{mot}(b), we show all possible subgraphs of sizes two, three, and four that can be described within three steps of a random walker thanks to the presence of the sink node. To avoid confusion with the motifs in Fig.~\ref{mot}(a), we use roman number to indicate the eight patterns in Fig.~\ref{mot}(b).It is worth to note that the introduction of the sink node allows to compute the frequency of paths whose length is smaller than the number of the maximum possible steps of the random walker. In our specific case, weighted subgraphs I, II, and III in Figure~\ref{mot}(b) would not be detectable without the sink node strategy. 
One could be interested in describing only motifs ending up in the sink node, but their number is included in our counting. Indeed, the group of all motifs observable within three steps of a random walker ending up in the sink node is a subset of the group of motifs observable within three steps of a random walker ending up in any node of the network. The choice of focusing on this subset of weighted subgraphs depends on the scope of the study. It is possible to compute network subgraphs of any dimension, increasing the random walker's maximum number of possible steps. Of course, this requires high computational costs and increases the complexity of the analytical computations.
We want to stress the critical role played by the sink node. It allows to (i) take into account the heterogeneity of incoming and outgoing fluxes in the definition of weighted motifs and (ii) to consider the emergence of those configurations as mutually exclusive events. Indeed, the identical network binary topologies can be characterized by very different weighted subpatterns according to the distribution of the weights. This approach allows comparing the eight configurations in Fig.~\ref{mot}(b) in terms of relative frequencies---as independent events---showing the composition of the network structure in terms of subpatterns organization and offering insights about its underlying mechanisms. 

\subsection{Analytical computations}

\indent The small number of allowed steps  for the random walker (3), permits to analytically compute the occurrences of the weighted motifs in Fig.~\ref{mot}(b). 

Given the transition probability matrix $M$ we can compute its $k$-th power
\begin{equation}
M^k = \Bigl(m_{ij,k}\Bigr)^{k=1,2,\dots}_{1 \le i,j \le N}
\end{equation} 

We know that $m_{ij,k}$ represents the probability that a random walker moves from node $i$ to node $j$ in exactly $k$ steps. In our case, we will only consider $k \in \{1,2,3\}$, but the following results generalize straightforwardly. 

Let us introduce the vector \textit{sink} such that
\begin{equation} 
{\rm sink}_i = 
\begin{cases}
(s_i^{\rm in} - s_i^{\rm out}) s_i^{\rm in} & \text{ if } s_i^{\rm in} -s_i^{\rm out} >0\\
0 & \text{otherwise} 
\end{cases}
\end{equation}

\noindent $\forall i \in \{1,\dots,N\}$. 
The theoretical probability of observing the motifs of Fig.~\ref{mot}(b) will be given by:
\begin{subequations}
\begin{align}
P_{\rm I} & = \Bigl( \sum\limits_{i,j=1}^N m_{ij}\; {\rm sink}_j\Bigr) / N,\\
P_{\rm II} & = \Bigl(\sum\limits_{i=1}^N m_{ii,2} \; {\rm sink}_i \Bigr)/N,\\
P_{\rm III} & = \Bigl( \sum_{\substack{i,j=1\\ j \neq i}}^N m_{ij,2}\;{\rm sink}_j\Bigr) / N,\\
P_{\rm IV} & = \Bigl( \sum\limits_{i,j=1}^N m_{ij}m_{ij}m_{ji}\Bigr) / N,\\
P_{\rm V} & = \Bigl(\sum\limits_{i=1}^N m_{ii,3} \Bigr)/N,\\
P_{\rm VI} & = \Bigl( \sum\limits_{i,j=1}^N m_{ii,2}m_{ij}\Bigr) / N - P_{\rm IV},\\
P_{\rm VII} & = \Bigl( \sum\limits_{i,j=1}^N m_{ii,2}m_{ji}\Bigr) / N - P_{\rm IV},\\
P_{\rm VIII} & = \Bigl(\sum_{\substack{i,j=1\\ j \neq i}}^N m_{ij,3}\;{\rm sink}_j\Bigr) / N - P_{\rm IV} - P_{\rm VI} -P_{\rm VII}.
\end{align}
\end{subequations}

\subsection{Simulations}

When the maximum number of steps of the random walker is greater than three, the number of weighted patterns exponentially increases and the analytical computation becomes harder. Therefore, we can compute the occurrences of weighted motifs through simulations (a Matlab package is available at~\cite{Matlab}). In this case, we performed $10^7$ Monte Carlo simulations for each network. At each iteration, the random walker started from a new node drawn uniformly at random from the set of all nodes. Then, we computed the frequency of each motif in Fig.~\ref{mot} (b).

Independently on how we computed the frequencies of each weighted motifs in the real case (analytically or computationally), we need to compare them with occurrences in an ensemble of network randomizations.  The choice of a proper null model to test the statistical significance of network properties is a crucial step, as the selection of the local constraints can affect the outcome~\cite{artzy2004comment,thusresponse}. We have tested three different null models, based on the maximum-entropy approach~\cite{squartini2011analytical,squartini2015unbiased}, with different types of local constraints: The Binary Directed Configuration Model (BDCM) with pure binary constraints, in/out node degrees, and reshuffled weights; (ii) the Weighted Directed Configuration Model (WDCM) with pure weighted constraints, in/out node strengths; (iii) the Enhanced Directed Configuration Model (EDCM) with binary and weighted constraints, in/out node degrees and strengths. Our study focuses on the third model corresponding to the more conservative null model as both degree and strength are kept fixed. Indeed, we think that for unbalanced weighted networks, both the number of incoming/outcoming links and the related total entering/exiting flows are fundamental to characterize the system~\cite{mastrandrea2014enhanced}. Moreover, as stated in Ref.~\cite{mastrandrea2014enhanced} the two measures, even if quantitatively related, are mutually irreducible according to the information contained. In the section \textit{Changing null model} we report the results related to the other two null models, putting in evidence similarities and differences in the significance profile of the weighted subgraphs.

For each null model, we generated 1000 randomizations and computed the z-scores to test the significance of each motif:
\begin{equation}
z_i = \frac{P^{\rm obs}_i - \mu ^{\rm rand}_i}{\sigma^{\rm rand}_i} \quad i = 1,2,\dots,8
\end{equation}

\noindent where $\mu^{\rm rand}_i$ and $\sigma^{\rm rand}_i$ are, respectively, the mean and the standard deviation of the motif frequency computed over the ensemble of randomizations.

\subsection{Examples}

\begin{figure}[!ht]
\centering
{\includegraphics[width=0.83\columnwidth]{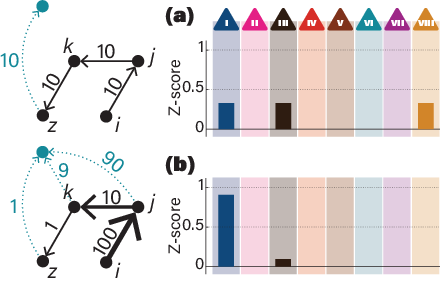}}
\caption{{\bf Example of weighted motifs composition for a simple network.} Two toy models having same binary topology and different weights distribution: homogeneous (top); heterogeneous (bottom).}
\label{example}
\end{figure}

In figure \ref{example} we show two toy models sharing the same topology, but with different weights distributions: homogeneous (top), heterogeneous (bottom). Without introducing the sink node, the transition probability matrices associated to the graphs are exactly the same. Therefore we are not able to differentiate between the two cases. On the contrary, the sink node, compensating for the excess of nodes in strength, favors the emergence of the most likely configuration in the heterogeneous case at the expense of the other two possible weighted motifs (the open path of order two and four, respectively). 

\begin{figure}[!ht]
\centering
{\includegraphics[width=\columnwidth]{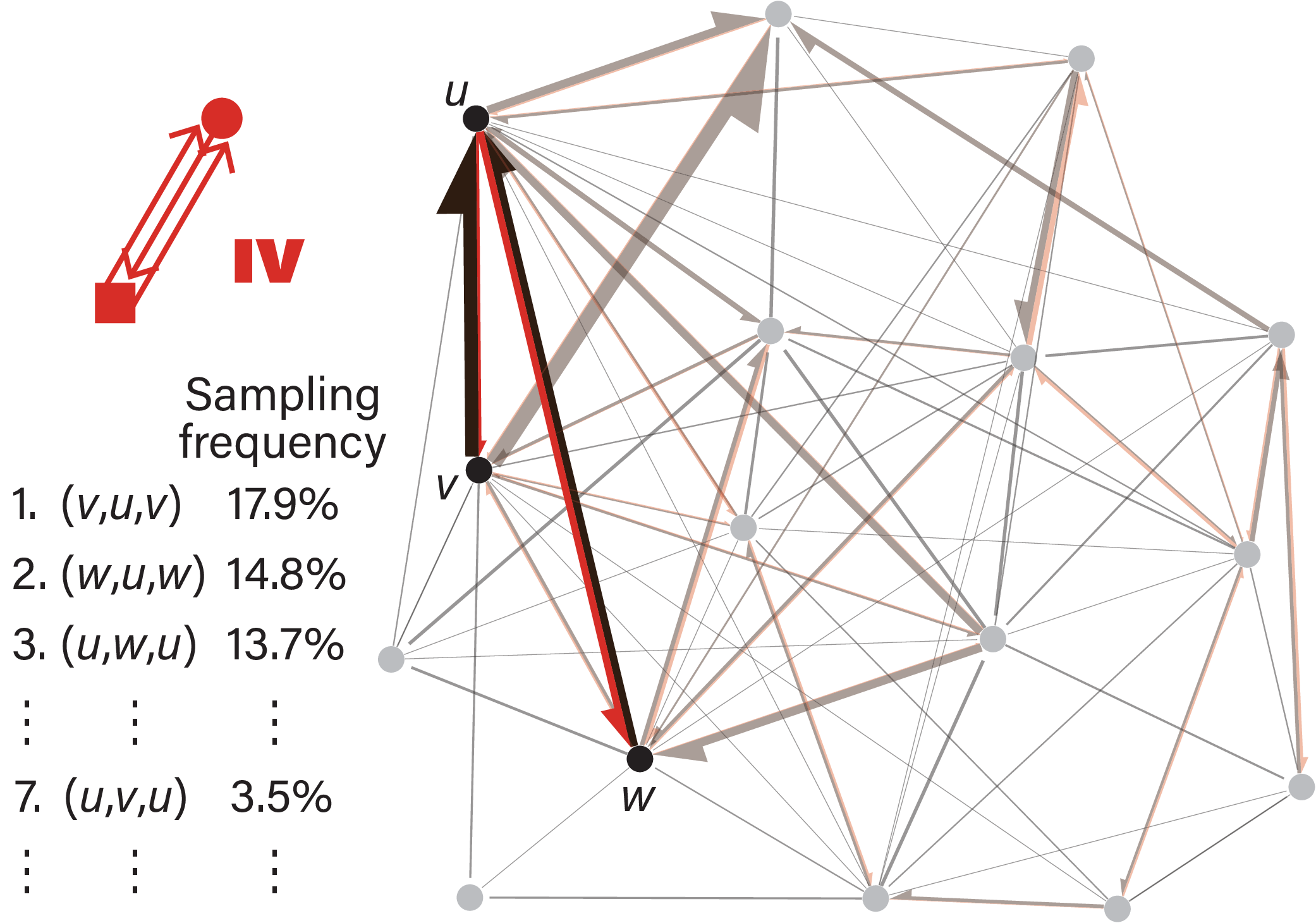}}
\caption{{\bf The most frequently sampled motifs in an empirical network of grooming in a troop of Rhesus macaques.} We show the most overrepresented weighted network motif IV, z-score 9.4, and some of its most frequently sampled instances involving three nodes---$u$, $v$, and $w$. The second most overrepresented motif in this example is VI (z-score 6.9).}
\label{macaques}
\end{figure}

Figure~\ref{macaques} reports a real-world example---a directed network of grooming among a troop of Rhesus macaques~\cite{macaques}. The most prominent weighted motif is IV of Fig.~\ref{mot}(b), which is a motif involving only two nodes. This suggests a relatively strong degree of reciprocity---grooming does not follow an absolute hierarchy in the troop. Note that the individual instances of network motifs are not sampled with the same frequency in weighted networks.

\section{Results}

\subsection{Weighted building blocks}

Figure~\ref{bar} shows the frequency the of weighted subgraphs of Fig.~\ref{mot}(b) in different real networks (see Appendix A for a dataset description). Within three steps, the analytical computations are still possible (see \textit{Methods}); for longer paths, they become too complicated, counting is only possible via numerical simulations. The frequencies are obtained normalizing motifs occurrences with respect to the total frequency of the eight motifs under study (patterns ending in the sink node in one step are neglected). At first sight, the bar charts in Fig.~\ref{bar} can give an immediate and intuitive idea about the different nature of the networks under study, shading light on their structural features. However, we need a null model to test the statistical significance of weighted patterns occurrence and classify networks accordingly. 

\begin{figure}
\centering{\includegraphics[width=\columnwidth]{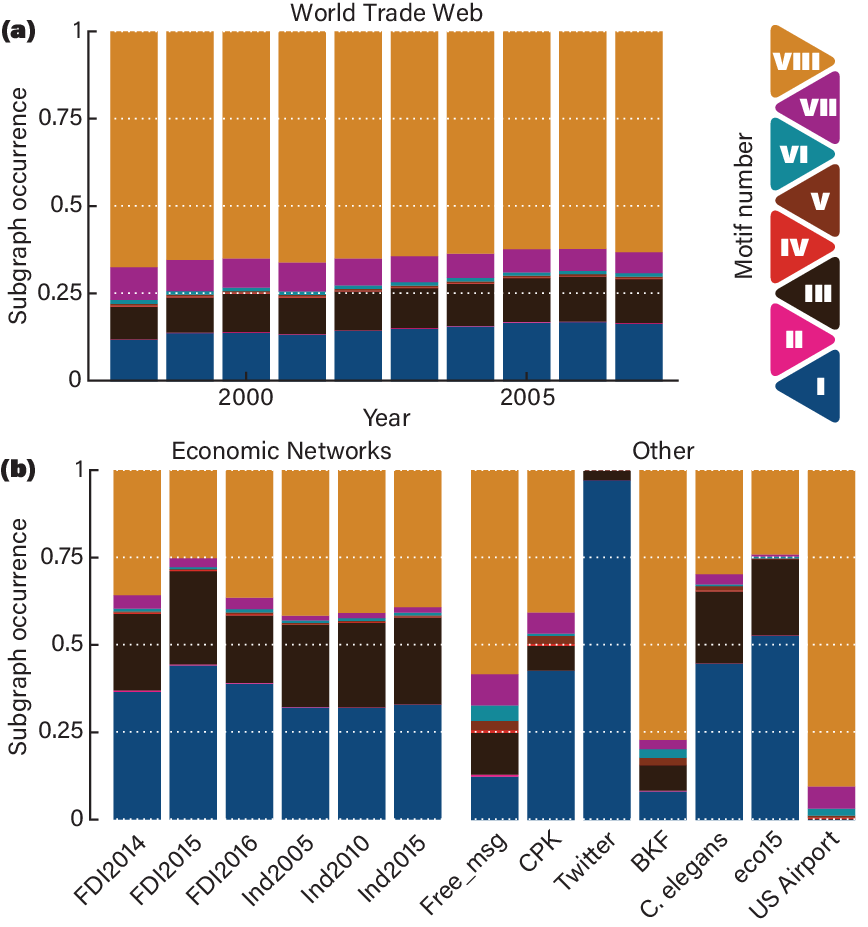}}
\caption{{\bf Bar-chart of weighted motifs composition for different networks.} normalized occurrence of weighted motifs in Fig.~\ref{mot} shown for several networks: (a) The World Trade Web (WTW) from 1998 to 2007; (b) the Foreign Direct Investment (FDI) network in 2014 and 2016; UK Input-Output table in 2005, 2010 and 2015; (c) two social networks: Freeman messages and Bk-office; two foodwebs: Mas Palomas basin and; Twitter network (highly reduced); \textit{C. elegans} neuronal network; US top 200 airports network.
\label{bar}}
\end{figure}

\begin{figure}
\centering{\includegraphics[width=\columnwidth]{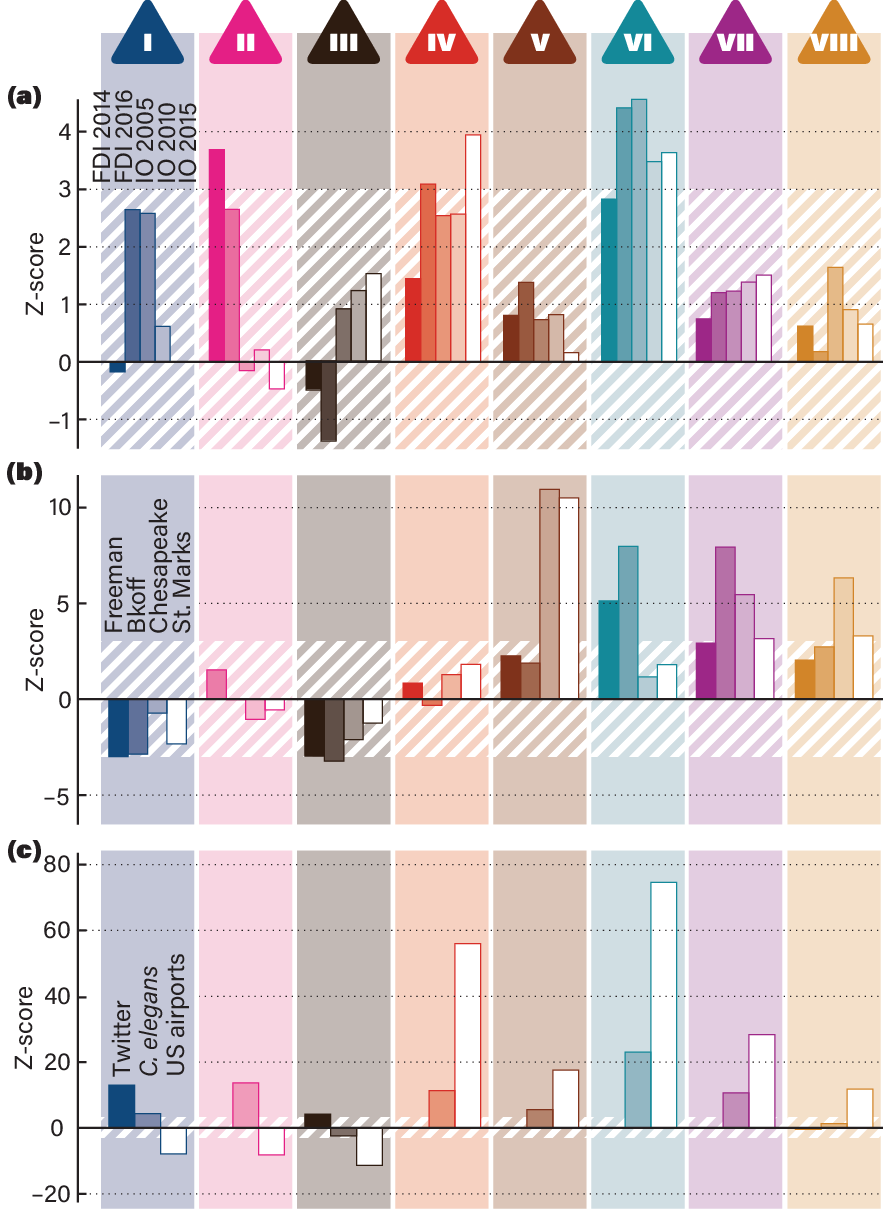}}
\caption{{\bf Z-score of weighted motifs under the EDCM null model.} Panel (a) shows the Foreign Direct Investment (FDI) network in 2014 and 2016 and the UK Input-Output table in 2005, 2010 and 2015. Panel (b) gives two social networks, Freeman messages and Bk-office and two foodwebs, Mas Palomas basin and St.\ Marks. Panel (c) displays a Twitter network (highly reduced), \textit{C. elegans} neuronal network and the US top 200 airports network. (Diagonal lines $z=\pm3$) 
\label{zscore_bar}}
\end{figure}

\begin{figure*}
\centering{\includegraphics[width=\textwidth]{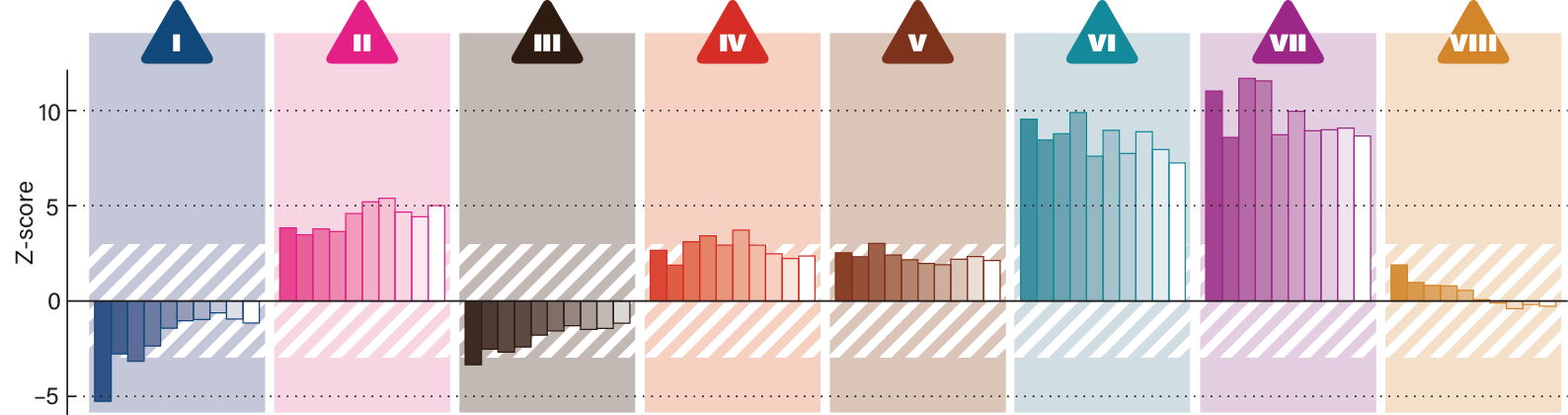}}
\caption{{\bf Z-score of weighted motifs under the EDCM null model.} The World Trade Web (WTW) from 1998 to 2007. (Diagonal lines indicate $z=\pm3$.)
\label{wtw_zscore_bar}}
\end{figure*}

The z-score profiles of the eight weighted motifs correctly identify systems belonging to the same field. Indeed, economic, social and ecological networks shows interesting similarities in the significant occurrence of some weighted motifs (fig.~\ref{zscore_bar} (a)-(b) and ~\ref{wtw_zscore_bar)}). This outcome appears in line with our initial scope to use the defined subgraphs as ``simple weighted building blocks'' extending the work by Milo \textit{et al.}~\cite{milo2002network} to the weighted case. Furthermore, it is possible to identify specific weighted subgraphs characterizing networks of different nature informing on the existence of common underlying functioning mechanisms.

We found clear evidence of weighted motif VI being overrepresented with respect to the null model in all networks except foodwebs and Twitter. The introduction of the sink node allows looking at the occurrences of motifs as independent events. This means that despite the similarities concerning motif VI, relevant differences can emerge. Indeed, motif VI contains motif II, which is also overrepresented in the economic and neuronal networks but underrepresented in the airport networks. Since we are considering random walks of three steps maximum, if motif II is overrepresented, it means that the walker usually ends up in the sink after two steps. Nevertheless, in the economic networks and the neural one, we observe that motif VI and IV are also overrepresented, implying that the third step of the random walker goes somewhere else from the sink node. On the contrary, for the airport data, the underrepresentation of motif II is in line with the overrepresentation of motif IV and VI. In other words, in the airport case, we observe a smaller number of motif II than in the random case. This means that most of the time, the walker comes back to the previous node (motif IV) or goes to a third one (motif VI). Therefore, the abundance of motif IV and VI is partly due to the scarcity of motif II. This outcome explains the nature of the top 200 US airports (remembering that we are considering only walks of length three)---two hubs tend to be connected by more than one flight per day (motif IV), while from one hub there are flights towards different destinations (motif VI). It could be interesting to explore more extended patterns to know more about the US airport network architecture. Economic and neuronal networks are different because the abundance of motifs II, IV, and VI implies that the three patterns coexist with motif VI more overrepresented than the others. Hence, the network is featured by an organization with three different configurations where none serves as a base for the others. Still, all are equally important for describing the functioning mechanism of economic exchange or neurons interactions in the nematode.

Motif VII looks similar to VI, but the different position of the starting node (central in motif VI, peripheral in motif VII) changes its interpretation. Indeed, the occurrence of motif VII is closely related to motifs III, V, and VIII. First of all, differently from motif VI, motif VII is not significant for FDI and IO networks, while it is for ecological networks. In all cases, we observe an abundance of closed cycles and a scarcity of open cycles of order three. Therefore, most of the time, the walker does not end up in the sink node after moving from node $i$ to $j$ to $k$, but it comes back to node $j$ or node $i$ forming a closed triangle (motif V). This is particularly evident for the US airports, less for the other networks. Different is the case of the ecological networks, where the occurrence of motif V is much higher than that of motif VII. Social networks (except Twitter, for its specific nature) tend to form more closed triangles than in the randomized networks. This appears to be a peculiarity of those networks with respect to the other systems analyzed here. Also, the open cycle of order four (motif VIII) seems to be overrepresented in all networks except in the economic ones. Therefore, patterns involving more than three nodes do not seem to characterize the economic exchange (WTW, Ind, FDI), at least in this context. 

The possibility to have the relative frequencies of weighted subgraphs in Fig.~\ref{mot}(b) and the related z-score profiles for each network under study offers insights about the organization of systems and their functioning, allowing the emergences of similarities and differences independently on the field they belong to. In this sense, our random walk patterns can be viewed as ``building blocks'' in the same way as statical motifs~\cite{milo2002network}.

\subsection{Temporal Dynamics}

Over-time changes of weighted subgraphs frequencies can reveal interesting hints about the structural reorganization of networks in time due to exogenous factors. In Ref.~\cite{squartini2013early} Squartini \textit{et al.}\ studied the Dutch interbank network focusing on binary motifs. They showed how the temporal trend of specific binary patterns worked as an early signal of the 2008 economic crisis. In the same spirit, Ref.~\cite{saracco2016detecting} studied the WTW searching for indications of the 2007/2008 financial crisis in the network organization, now looking at the bipartite network of countries and products and related \textit{ad hoc} binary motifs. They found that the WTW became increasingly compatible with a network where correlations between countries and products were progressively lost starting from 2003.

This section reports three examples of a temporal dynamic study for specific countries in the WTW. We used a combination of two different datasets to have a longer reference period, including some relevant economic/historical/political events (see \textit{Dataset}). We computed the occurrence of the weighted motif when the random walker started from the same node at each iteration of the Monte Carlo simulation. In this way, we considered only weighted motifs having the specific country as the initial node of all weighted motifs. The temporal dynamics of these patterns could be in a cause-and-effect relationship with historical and political events.

\begin{figure}
\centering
{\includegraphics[width=\columnwidth]{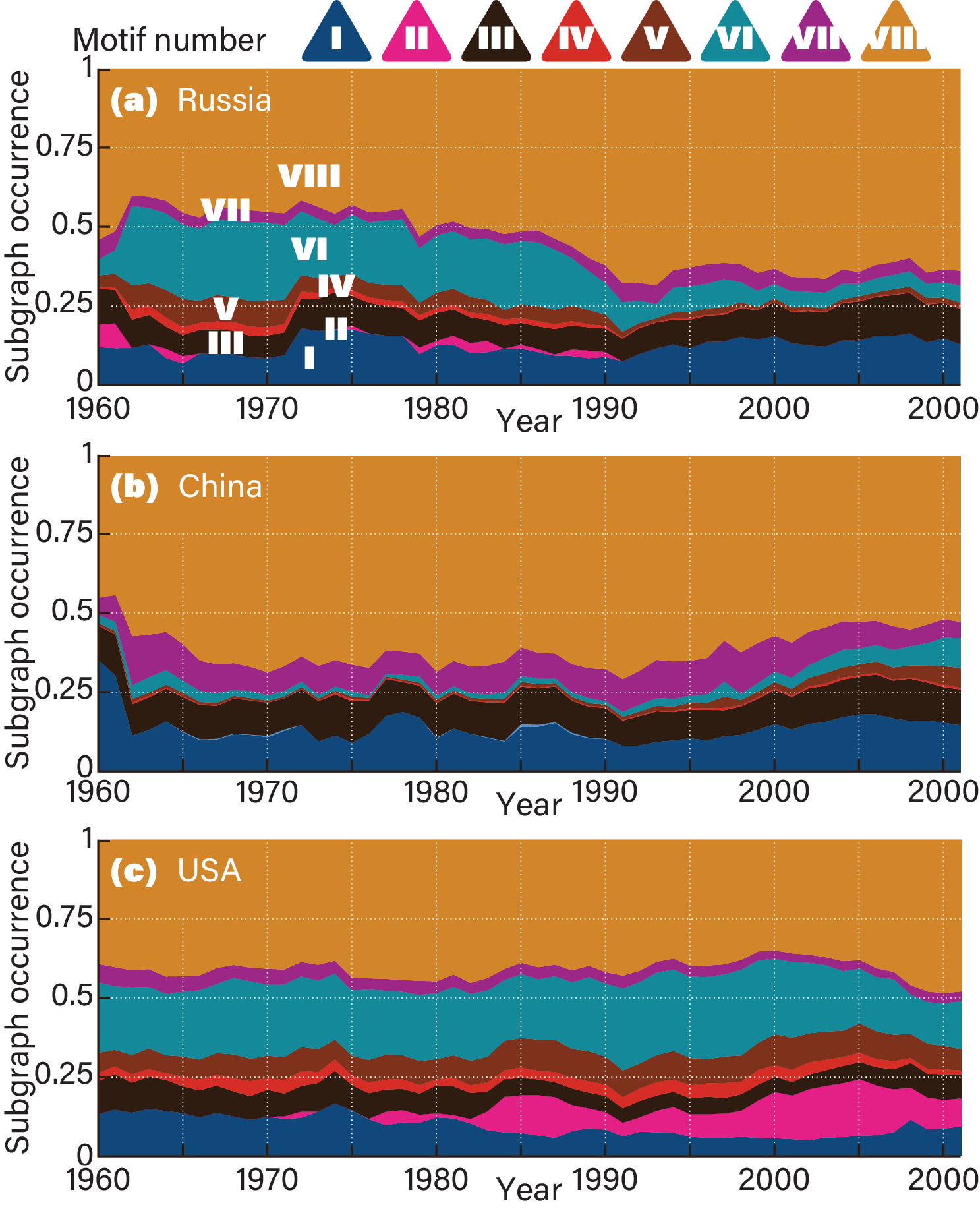}}
\caption{{\bf Temporal dynamics of weighted motifs.} Weighted motifs computed for the WTW data from 1960 to 2011 considering a random walker starting from the same node/country: (a) Russia, (b) China, (c) USA. 
\label{trend}}
\end{figure}

Although almost one-quarter of the value of the WTW (World Trade Web) is cyclical---i.e., it returns to the origin in the form of accrued value through international production chains~\cite{picciolo2017crude}---the most recursive weighted motifs are open paths (motifs I, II, VII, and VII in Fig.~\ref{mot}(b)). It is thus likely that global value chains take longer than three steps to complete their cycles. By taking the perspective of countries (Fig.~\ref{trend}), it is possible, for example, to discern how the globalization process, which began in the late 1990s, shaped the embedding in global value chains for China, Russia, and the USA. While Russia saw a collapse in the length and complexity of patterns, with the emergence of motif VIII (long open paths) at the expense of the others (suggesting that Russia became mainly a supplier of raw materials), China witnessed the significant rising of motifs V and VI, hinting to its increasing involvement in global value chains, as confirmed by a previous analysis on clustering in the flows of embodied value-added~\cite{amador2018s}. Pendant to the rising of China as a hub of complex motifs, the USA saw an increase in motifs II and VIII, highlighting, on the one hand, their increased marginalization in the global value chains, on the other their tighter bilateral ties (mainly with China).

This section illustrated one possibility to use the proposed approach to explore the underlying motif structure and its possible interconnection with exogenous events. 

\subsubsection{An application: USA trade balance }

\begin{figure}
\centering
{\includegraphics[width=\columnwidth]{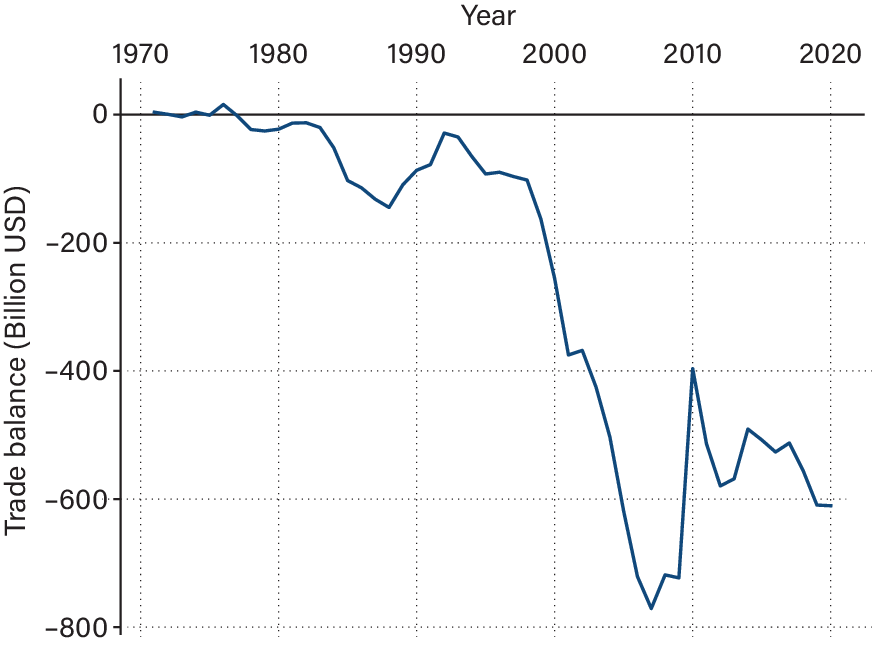}}
\caption{{\bf USA balance of trade.} Source: www.tradingeconomics.com }
\label{app}
\end{figure}

The balance of trade of a country in the WTW is the difference between its total export and import. A negative balance means that the country is importing more than what it is exporting. Figure~\ref{app} shows the balance of trade of USA from 1960 to 2013~\cite{balance}. Its trend shows interesting relation with the temporal dynamic of the weighted motif II in Fig.~\ref{trend}(c). If the negative balance of country $i$ increases, there is a high probability for a random walker starting from node $i$ to come back to the same node and end up in the sink node. In other words, the likelihood to observe pattern II is higher than that of detecting subgraphs IV or VI. It is worth noticing that weighted motifs I, II, and III are never observable in a balanced network.

\subsection{Changing null model}

\begin{figure*}
\centering
{\includegraphics[width=0.85\textwidth]{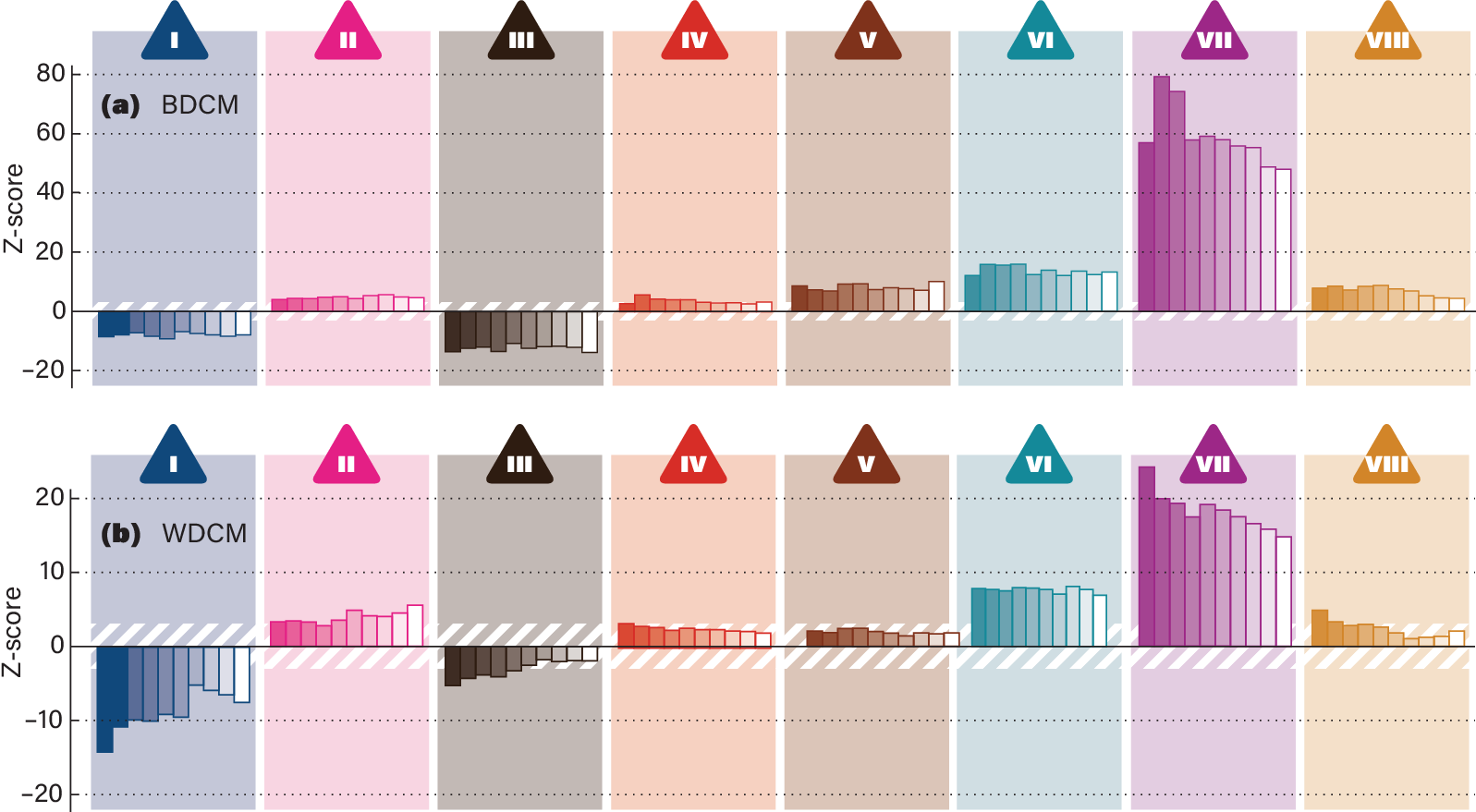}}
\caption{{\bf Z-score of weighted motifs for the WTW data.} The z-scores are computed under different null models---BDCM and WDCM (for EDCM, see Fig.~\ref{zscore_bar}). Diagonal lines indicate z=$\pm3$. For every weighted motif, the bars are ordered from earlier (left) to later (right). One bar represents one year from 1998 to 2007.
\label{zscore3}}
\end{figure*}

This section highlights the fundamental role played by the null model in testing the significant occurrence of network weighted subpatterns. Moreover, the comparison between models using different local constraints can shed light on the structural organization of the network under study. In Fig.~\ref{zscore3} we show the z-score of the eight motifs in Fig.~\ref{mot}(b) for the WTW computed with respect to three different ensembles of randomizations: (i) BDCM; (ii) WDCM and (iii) EDCM. The comparison reveals a similar profile for all motifs: the z-score value decreases (in absolute term) as the weighted local constraint is taken into account, particularly for motifs VII and III, but also V and VI. More specifically, this means that when node strength is introduced as a local constraint for the network randomization, the overestimation of weighted motif VII drastically reduces, even if it is still significant. In contrast, the occurrence of motif III becomes not significant. Given the definition of the transition probability (see Methods), this result suggests that node strength plays a major role for the information carried by node degree in shaping the structural organization of the WTW architecture and determining the significant occurrence of motifs. This feature is not common to all networks in the sample.

\begin{figure}
\centering
{\includegraphics[width=\columnwidth]{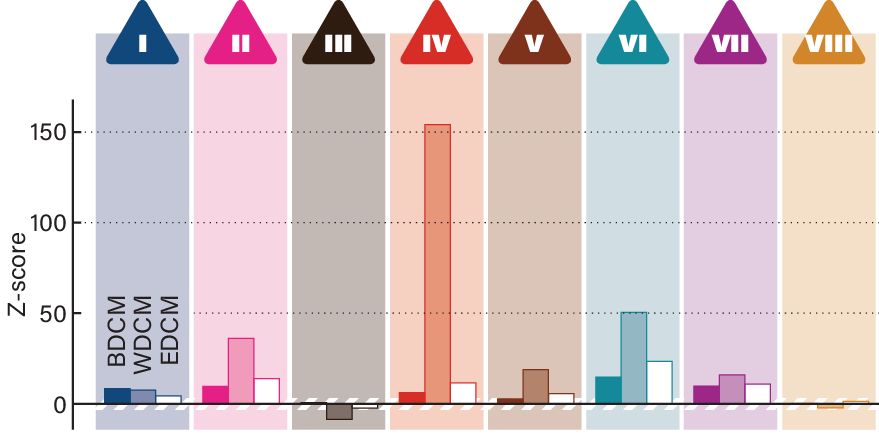}}
\caption{{\bf Z-score of weighted motifs for the \textit{C. elegans} neuronal network.} The z-scores are computed under different null models: BDCM, WDCM and EDCM. (Diagonal lines indicate $z=\pm3$)
\label{zscore4}}
\end{figure}

In Fig.~\ref{zscore4} we show the significance profile of the eight motifs for the nematode network \textit{C. elegans}. In this case, we observe that: (i) there is a similarity in the significance profile of the eight motifs for the three null models; (ii) the z-score is very high under the WDCM and shrinks when the information about the degree is introduced; (iii) the node degree as a constraint is more informative when considered alone than when also the node strength is added (iv) the significance ranking of motifs changes according to the null model used (IV, VI, II, VII under WDCM, VI, II, IV, V under EDCM and BDCM); (v) motif III is underrepresented with respect to the WDCM, but become not significant under EDCM and BDCM. All these results suggest that for the nematode's synaptic network, the role of strengths is not relevant as for the WTW in describing some network subpatterns, and consequently, their formation mechanism. On the contrary, the information about node degree, even if not sufficient to explain such patterns, reduced the discrepancy between the observed and the randomized occurrences~\cite{squartini2013reciprocity}.

We single out these two results to show the importance of selecting a proper null model and interpreting the results accordingly. It would be interesting to investigate more the role of each local constraint in shaping the structure of empirical networks in different fields, but for the moment, this is out of the scope of this paper. The key general result is that---analogous to no-free lunch theorems~\cite{wolpert1997no}---no null model does \lq better \rq{} than another. One needs to consider the specifics of the problem and carefully draw conclusions considering these. 

\section{Discussion}

In this paper, we have introduced a novel methodology based on a random walker to study the occurrence of weighted subgraphs in networks. In line with work of Milo \textit{et al.}~\cite{milo2002network}, our goal consists in characterizing networks belonging to different fields according to the existence of ``weighted motifs'' as a sort of building blocks. They could help in unraveling the underlying functioning mechanism and the evolutionary processes that brought to network formation. The introduction of weights can alter the outcomes of the binary case. At present, the vital role played by weights in providing a complete characterization of a complex network is well known (especially when diffusion processes, propagation mechanism, network resilience, and robustness to external shocks are studied)~\cite{newman2004analysis,barrat2004architecture,fagiolo2008topological}.

A consensus definition of weighted motifs has so far been missing. Indeed, it has been pointed out that adapting unweighted methods to weighted networks is not always straightforward---the definition of the weighted clustering coefficient is a good example where there are various definitions, seemingly all well-motivated~\cite{barrat2004architecture,HOLME2007821,fagiolo2007clustering,saramaki2007generalizations,fagiolo2008topological,opsahl2009clustering}. According to our knowledge, only one paper (Ref.~\cite{onnela2005intensity}) proposed a generalization of the concept of motifs to the weighted case introducing two measures \textit{intensity} and \textit{coherence} complementing the binary motifs. However, they do not expand on these two quantities by characterizing the network's properties but only show the example of the \textit{E. Coli} network comparing some binary and weighted motifs. 

We base our approach on finite length paths traced by a random walker starting from a random node and allowed to move for a limited number of steps according to established transition probabilities. The novelty of the methodology consists in balancing the network introducing a \textit{sink} node for compensating the excess of node in-strength before computing the transition matrix.

The introduction of the sink node allows distinguishing between networks having the same binary topology but different weights distribution. The addition of the sink node puts in evidence the role played by the weights heterogeneity in determining the increase (decrease) of occurrence probabilities of some weighted motifs at the expense of others. Indeed, all weighted subgraphs smaller than the maximum possible number of steps can be considered mutually exclusive events whose total occurrence probability sum up to one. In this way, we have information about the system's composition in weighted patterns in terms of shares. In our specific case, we consider a maximum of three steps that allow us to define all configurations in Fig.~\ref{mot}(b). 

In the spirit of Ref.~\cite{milo2002network}, we consider empirical networks of different origin to be characterized  by their organization in significant sub-patterns. We first show the network composition according to the motifs of Fig.~\ref{mot}(b). Figure~\ref{bar} reveals the economic networks showing high stability over time (Fig.~\ref{bar}(a)-(b)), the social networks, and the airport web exhibiting a great abundance of open paths of order four (motif VIII), while the open paths of order three is absent in the latter. Although we can deduce the different network organizations in sub-patterns from these results, they do not say anything about their significant occurrences. Indeed, the abundance of a specific motif could be simply ascribed to a statistical effect. Due, for example, to network size or \textit{connectance} (the fraction of edges existing among all node-pairs). For this reason, it is necessary to compare their frequency in real networks with specific null models. Since we are considering directed weighted networks, we selected a null model fixing the two main local properties: nodes degree and strength. The choice of the null model is relevant because it strongly depends on the type of the network and the sub-patterns under examination~\cite{artzy2004comment,thusresponse}. We have shown in the section ``Changing null model'' how the outcome of the analysis is sensitive to the choice of the null model and, consequently, the interpretation in the perspective of network underlying formation mechanism and evolution. 

As a first relevant result, we found that networks belonging to the same field have close motifs significance profiles: economic networks (Fig.~\ref{zscore_bar}(a) and ~\ref{wtw_zscore_bar}); social networks (Fig. ~\ref{zscore_bar}(b)); ecological networks (Fig.~\ref{zscore_bar}(b)). The comparison between the three different economic datasets considered: the international trade network, the FDI network, and the IO tables, are worth noting. Indeed, the FDI networks show a significance profile of motifs lying between the WTW and the IO tables, revealing differences and similarities of the Mergers \& Acquisitions networks with the two economic systems (for further details about the three datasets, see the dedicated section).

The significance profile of the weighted motifs of each network reveals something about its underlying organization and evolutionary mechanism of formation. For example, two very different networks, as the top 200 US airports and the neuronal system of a nematode, share a very similar significance profile for motifs IV, V, VI , VII, while patterns I and II appear overrepresented in the \textit{C. elegans} network and underrepresented in the airport network. Bearing in mind the rules of the random walker and the nature of the system, we can deduce that it is very rare to find one-way only flights (motif I) in the airport network or just two-way flights between pairs of nodes (motif II). At the same time, it is relatively likely to observe motifs IV or VI. On the contrary, in the neural network, it is possible to observe an abundance of the pattern I and II together with IV or VI. Of course, we should search the reason for such differences in the different roles played by neurons and airports in their respective systems and the specific meaning of their connecting edges. This is one of the possible applications of our study, especially if higher-order paths are considered.

Our approach also allows us to identify specific subpatterns characterizing almost all the different systems, revealing critical functional features. This is, for example, the case of motif VI, overrepresented---with respect to the null model---in all networks except foodwebs and Twitter. Furthermore, it is possible to focus on selected nodes, study the over time variation of the weighted motifs composition, and relate them with exogenous shocks as economic, political, and social events. We have shown three examples from the WTW dataset concerning countries that experienced different historical occurrences with consequent different weighted patterns temporal frequencies (Fig.~\ref{trend}).

In conclusion, we have presented a novel approach to identify network subgraphs in directed weighted networks. We have shown that the significant abundance (scarcity) of such patterns compared to a properly chosen null model can shed light on the nature of the systems, their functioning, and the evolution mechanism that generated them. By these means, we can classify networks according to some universal rules, beyond field-specific knowledge and acquisitions. Of course, the analysis's limitation consists of the small number of steps allowed to the random walker. However, as previously explained, this depends on the super-exponential increase of complexity as the number of maximum steps increases. Moreover, studying a broader collection of empirical networks would be rewarding, including networks belonging to other fields not included here. Our choice strictly depended on network sizes and the time consumption for generating the ensemble of randomization. In the future, we hope to overcome this limitation, improving the performance of our maximum-entropy approach to creating the ensemble of null models. Lastly, we think it would be fascinating to explore more of the architecture of some networks by simultaneously looking at the occurrence of binary and weighted subgraphs in relation to the nature of the systems, their formative mechanisms, the meaning, and the roles of its components (node and links) together with other higher-order topological properties and exogenous events.

\section*{Acknowledgments}

RM acknowledges support from the Italian ``Programma di Attivit\`a Integrata'' (PAI) project ``PROsociality COgnition and Peer Effects'' (PRO.CO.P.E.), funded by IMT School for Advanced Studies Lucca.
PH acknowledges support from JSPS KAKENHI (grant no.\ JP 18H01655).

\section*{Appendix: Dataset}

\subsubsection{The World Trade Web}We used international trade data provided by the BACI database for building the network analyzed in Figs~\ref{bar}(a).
They represent the monetary imports and exports among world countries (in millions of current US dollars, bilaterally harmonized). The original data was provided by the United Nations Statistical Division (COMTRADE database). BACI is constructed using a procedure that reconciles the declarations of the exporter and the importer. Further details can be found in the CEPII Working Paper~\cite{gaulier2010baci}. All networks have $N = 208$ nodes and average density $\sim 0.35$. For the temporal dynamics studies we combined the BACI dataset from 1998 to 2011 and the GLEDTISCH dataset from 1960 to 1997~\cite{gleditsch2002expanded} to have a longer observation period.

\subsubsection{Foreign Direct Investment}
The data are extracted from Worldwide Mergers, Acquisitions, and Alliances Databases SDC Platinum (Thomson Reuters), financial databases providing information on global transactions from 1985 to 2010. Here we considered only two years: 2014, 2016. They represent the monetary investments from a target country to an acquirer one (data are aggregated over all sectors involved in the investing process between two countries). Most recorded transactions refer to domestic Mergers \& Acquisitions (M\&A) activity ($\sim 74\%$). The nominal monthly M\&As inflows and outflows are deflated using the Industrial Production Index provided by the US Bureau of Labor Statistics~\cite{defl}. The resulting directed weighted networks consist of rows indicating investors and columns standing for receivers/targets. They are described in great detail in Ref.~\cite{duenas2017spatio}. The network for 2014 has $N = 151$ nodes and connectance equal to $0.05$; for 2014 $N = 146$ nodes and connectance equal to $0.05$. 

\subsubsection{Input-Output tables for the UK}
This dataset comes from the Input/Output (IO) table for UK~\cite{IOtable}. The IO table shows the interdependencies between different sectors of a national economy. The tables report the relation between input and output goods for different sectors. For example, the output of one sector can become the input of another one. 
All networks have $N=96$ nodes, and average connectance $\sim 0.75$.

\subsubsection{Freeman Message}
This dataset was collected in 1978 and contains three networks of researchers working on social network analysis. Here, we considered the frequency matrix of messages exchanged among 32 researchers that used an electronic communication tool~\cite{Freeman1979networkers}. The network has $N = 32$ nodes and connectance equal to $0.44$.

\subsubsection{Bkoff Social Network}
This is one of the ``classical'' social networks collected by Bernard and Killworth in bounded groups. It represents the network conversations frequency among the employees of a small business office, reported every 15 minutes (during two periods of four days each) by each worker (this explains the network asymmetry)
\cite{killworth1976informant}. The network has $N=40$ nodes and a very high connectance equal to $0.99$. 

\subsubsection{Chesapeake Foodweb}
This dataset represents the natural interconnection of food chains of what-eats-what in the ecological community of Chesapeake Bay Mesohaline~\cite{baird1989seasonal,Pajek}. The network has $N = 39$ nodes and connectance equal to $0.12$.
\subsubsection{Saint Marks Foodweb}
This dataset represents the natural interconnection of food chains of what-eats-what in the ecological community of the St. Marks River, Florida~\cite{Pajek}. The network has $N = 54$ nodes and connectance equal to $0.12$.

\subsubsection{Twitter}
Twitter data comes from the Stanford Network Analysis Project
(SNAP) as part of the Stanford Large Network Dataset Collection~\cite{SNAP}.  The 
\textit{Higgs dataset} includes Twitter messages about the discovery of the Higgs boson announced on the 4th of July 2012. Tee observed periods ranges from the 1st to the 7th of July 2012.  The resulting retweeting network is directed and weighted. The dataset has been largely reduced to $\sim 400$ users for time-consumption problems connected to the randomization technique. The network has $N=424$ nodes and connectance equal to $0.002$.

\subsubsection{\textit{C. elegans} Network}
This dataset represents the neural network of the nematode \textit{Caenorhabditis elegans}. Two neurons are connected if at least one synapse or gap junction exists between them. The weight is the number of synapses and gap junctions~\cite{achacoso1991ay,watts1998collective}. The network has $N=297$ nodes and connectance equal to $0.03$.

\subsubsection{Top 200 US Airports}
This dataset represents the number of passengers flying between the top 500 US airports in 2010 downloaded from Bureau of Transportation Statistics (BTS) Transtats site, ties with weight zero (only cargo) and self-loops removed~\cite{Air}. The network has been reduced to include only the top 200 US airports, it has $N = 200$ nodes and connectance equal to $0.19$. \\

\bibliography{RW}

\bibliographystyle{apsrev}

\end{document}